\documentclass{Interspeech2024}

\usepackage{graphicx}
\usepackage{subfig}
\usepackage[noabbrev,nameinlink]{cleveref}
\usepackage{array}
\usepackage{multirow}
\usepackage{paralist}
\PassOptionsToPackage{hyphens}{url}
\usepackage{hyperref}
\usepackage{tipa}
\usepackage{siunitx}

\newcolumntype{P}[1]{>{\centering\arraybackslash}p{#1}}

\interspeechcameraready

\title{Towards objective and interpretable speech disorder assessment: a comparative analysis of CNN and transformer-based models}

\name[affiliation={1}]{Malo}{Maisonneuve}
\name[affiliation={1}]{Corinne}{Fredouille}
\name[affiliation={2}]{Muriel}{Lalain}
\name[affiliation={2}]{Alain}{Ghio}
\name[affiliation={3}]{Virginie}{Woisard}

\address{
  $^1$LIA, Avignon University, France\\
  $^2$Aix-Marseille Univ, CNRS, LPL, Aix-en-Provence, France\\
  $^3$LNPL, Toulouse University and Toulouse Hospital, Toulouse, France
}
\email{(malo.maisonneuve,corinne.fredouille)@univ-avignon.fr, muriel.lalain@univ-amu.fr}


\begin{document}

\maketitle

\begin{abstract}
  Head and Neck Cancers (HNC) significantly impact patients' ability to speak, affecting their quality of life. 
  Commonly used metrics for assessing pathological speech are subjective, prompting the need for automated and unbiased evaluation methods. 
  This study proposes a self-supervised Wav2Vec2-based model for phone classification with HNC patients, to enhance accuracy and improve the discrimination of phonetic features for subsequent interpretability purpose.
  The impact of pre-training datasets, model size, and fine-tuning datasets and parameters are explored.
  Evaluation on diverse corpora reveals the effectiveness of the Wav2Vec2 architecture, outperforming a CNN-based approach, used in  previous work.   
  Correlation with perceptual measures also affirms the model relevance for impaired speech analysis. 
  This work paves the way for better understanding of pathological speech with interpretable approaches for clinicians, by leveraging complex self-learnt speech representations.
\end{abstract}

\section{Introduction}

Head and Neck Cancers (HNCs) affect the upper respiratory and digestive tracts, including the oral cavity, pharynx, larynx, nasal cavity and salivary glands.
Treatment of this cancer, either by radio-therapy, chemotherapy and/or surgery, can significantly impact the speech of HNC patients.
Difficulties in communication with others negatively impacts the patients' quality of life.
It is essential to correctly assess their speech production, identifying its level of impairment and what makes it atypical, so that they can benefit from the best possible care during rehabilitation sessions or for evaluating the efficacy of a prosthesis following a surgical procedure.
Unfortunately, commonly used metrics such as severity or intelligibility measures are subjective, and prone to misjudgments, even by experts~\cite{woisard2021c2si}.
Furthermore, these metrics provide no information, other than a score, on the nature of the measured degradations.
This highlights the crucial need for an automated, objective method of speech assessment that offers not only accurate results but also interpretable insights into the specific speech characteristics affected.
Some papers investigated the assessment of HNCs patients' speech intelligibility and/or severity with relatively simple networks~\cite{maier2010,middag2014,laaridh18b,vaysse2021}, as well as using deep neural networks~\cite{bin2019,quintas2023,abderrazek2023}. 

Recently, self-supervised learning (SSL) models have shown to be successful in capturing phonetic patterns and various speech features. 
In~\cite{dieck22_interspeech}, authors have shown that Wav2Vec2 models~\cite{baevski2020} can learn certain phonetic concepts, and that they correctly preserve the manner and place of articulation. 
\cite{pasad2021,pasad2023} have also shown that phonetic information is contained in the representations built by these models.
Some research has been focused on using these models to automatically assess the speech severity level~{\cite{hernandez22_interspeech,yeo23icassp,javanmardi2024}}.
Other studies analysed how well diseases can be predicted by these models.
For instance, A. Favaro et al. \cite{favaro2023} compared interpretable speech features to embeddings produced by SSL models on predicting the presence of Parkinson's disease.
They showed that using embeddings provides better detection accuracies at the cost of losing the insight into speech and language deterioration given by interpretable features.
While being able to detect a disease and assess its severity is important, we believe it is as important to interpret the output of these models, in order to enhance trust that clinicians can have in these systems.
So far, only a limited number of studies focused on interpreting these model outputs.
In~\cite{ming2017}, a model was trained to predict dysarthric speech severity, by adding a bottleneck layer in a feed-forward fully-connected neural network to improve interpretability.
Indeed, they relied on transfer learning to learn clinically interpretable labels taken from the Darley classification~\cite{darley1969b}.
Their results show an accuracy improvement in dysarthria speech evaluation, along with justifications based on interpretable characteristics.
An extension of this work in~\cite{xu2023} avoided perceptual labels (which require costly annotation by experts, and is expensive in terms of recording amount required to train and evaluate models) by shaping the interpretable layer around four automatically-extracted acoustic features characterizing four dysarthria aspects, very close to the previous work.
By using a SHapley Additive exPlanations~\cite{lundberg2017}, they analysed the contribution of each acoustic feature on the final prediction.
Yeo et al. have also shown the efficacy of leveraging the \textit{Goodness of Pronunciation} metric, by using the pre-trained multi-layer convolutional neural network part -- frozen -- of the Wav2Vec2 model, followed by a phone classifier~\cite{yeo23interspeech}. 
Their approach shows the relative impact of each phone on the predicted severity score.
Another methodology was used in \cite{abderrazek2023} by using a convolutional neural network (CNN) for a phone classification task.
The NCD method (\textit{Neuro-Concept Detector}) along with the ANPS metric (\textit{Artificial Neuron-based Phonological Similarity}) proposed in this work enables predicted severity and intelligibility scores to be interpreted in terms of alterations in the speech produced by patients, by associating the behaviour of neurons from the classifier with the phonetic feature they detect.

The work presented in this paper aims to revisit the overall framework proposed in \cite{abderrazek2023} by replacing the CNN with an SSL model based on the complete Wav2Vec2 architecture.
This modification is anticipated to provide not only higher accuracy, but also to bring later on a more nuanced and richer description of the phonetic features of pathological speech.
In addition to the change of model architecture, this work will explore how pre-training corpora, model size, the fine-tuning stage, and datasets associated with this stage affect the choice of a publicly available pre-trained Wav2Vec2 model for phone classification.
This exploration is pivotal for optimizing the performance of the model across diverse datasets, contributing to the robustness and generalizability of our approach.
Thus, a detailed analysis of phone confusions is conducted, to ensure the generalization ability of our models on other datasets.
Subsequently, we analyse how phone classification accuracies correlate with perceptual measures obtained from domain experts on speech from patients who have benefited from care following a HNC.
This multidimensional evaluation approach will provide a comprehensive assessment of the proposed SSL model, shedding light on its potential for an accurate and clinically relevant pathological speech analysis,  required for Step 2 and 3 of the targeted overall framework \cite{abderrazek2023}.

\section{Corpora}\label{sec:corpus}

To train and evaluate our models during training, we relied on subsets of BREF~\cite{lamel1991bref} and Common Phone~\cite{klumpp2022cp}.
To test our models, we relied on BREF-Int and C2SI~\cite{woisard2021c2si}.

The \textbf{BREF} corpus is composed of 120 French speakers reading extracts from the newspaper \textit{Le Monde}.
Recordings took place in the 90s, with people recruited in the Paris area.
A phone-balanced subset of this corpus was created to ensure that the subsequently fine-tuned models are not showing bias towards a specific phone.
It is composed of more than 3 million frames of \qty{127}{\ms}, each aligned on a French phone or silence.
We also use the \textbf{BREF-Int} dataset, a phone-balanced subset of the BREF test dataset.
These datasets are identical to the ones used in~\cite{abderrazek2023}.

The \textbf{Common Phone} corpus is a gender-balanced, multilingual and phonetically-aligned corpus derived from Mozilla's Common Voice project.
Only French recordings were used in this work, including speech from Belgium, Tunisian and Canadian speakers, amongst others.
This dataset is also balanced in terms of phones and genders to ensure the phone-wise and gender-wise impartiality of the models.

The \textbf{C2SI} corpus consists of 87 patients who have been treated for oral or oropharyngeal cancer, as well as 41 healthy controls (HC).
Patients and HC were recorded at the IUCT Oncopole, in Toulouse, France.
Patients were faced with multiple tasks: sustained /a/, sentence reading, image description (DES), short text reading (LEC), pseudo-words production (DAP) as well as various other prosodic tasks.
Based on the image description task, six experts evaluated the severity (degree of alteration of the speech signal) and intelligibility of the patients' speech on a scale from 0 -- strong alterations -- to 10 -- perfect speech; the intelligibility being defined here as the ``the performance by a listener to recognize the words and/or the sounds of the speech produced by the speaker''~\cite{woisard2021c2si}.
Recordings from both tasks LEC and DAP are used for testing our models.
They will be referred here as C2SI-LEC and C2SI-DAP.
To match the selection that has been made in a previous work, we will test our models on 24 HC (amongst all available speakers, some of whom have not completed the two LEC and DAP tasks) all recorded under the same conditions.
We will compare our results with perceptual evaluations of 81 patients -- 6 patients were not evaluated by experts.

All the aforementioned corpora are phonetically aligned with 31 phones and silence.
These 31 phones include four archi-phones: /Ê/ = \{e, \textepsilon\}, /Û/ = \{\oe, \o\}, /Ô/ = \{o, \textopeno\}, and  /µ/ = \{\~\oe, \~\textepsilon\}.
The use of these archi-phones neutralizes true-mid vowel oppositions.
\Cref{tab:corpus} details the amount of phone-aligned frames used from each corpus, as well as the number of hours they represent.
Because some frames overlap each other, this number of hours is higher than the sum of recordings lengths.
However, it still reflects what the model sees as input.
\begin{table}[ht]
  \centering
  \caption{Usage, number of frames, and global duration of audio data for each corpus used in this work.}
    \scalebox{0.97}{
    \begin{tabular}{p{0.28\linewidth}P{0.25\linewidth}P{0.15\linewidth}P{0.11\linewidth}}
      \toprule
      Corpus & Usage & \#frames & \#hours\\
      \midrule
      BREF & train, validation & 3,118k & 110h \\
      Common Phone & train, validation & 236k & 8.3h \\
      BREF-Int & test & 85k & 3h \\
      C2SI-LEC (HC) & test & 43k & 1.5h\\
      C2SI-DAP (HC) & test & 73k & 2.5h\\
      \bottomrule
    \end{tabular}
  }
  \label{tab:corpus}
\end{table}

\vspace{-1em}

\section{Model architectures}\label{sec:models}

The CNN model has been previously trained for phone classification in~\cite{abderrazek2023} on the BREF dataset described in \cref{sec:corpus}.
It is composed of two convolution layers combined with maximum pooling layers.
The model input is a sliding window of 11 acoustic frames of \qty{20}{\ms}, \qty{10}{\ms}-spaced, where each frame is characterized by 40 Mel-Filter bank feature extracted from audio signal, along with their first and second derivatives.
The model thus has a \qty{120}{\ms}-window centred on the phone to be predicted.
Once the CNN has been applied, the output is then flattened before being fed to three fully connected layers, detailed below.
This model output is considered as the \textit{baseline} later in the paper.

Regarding the Wav2Vec2 architecture, models from LeBenchmark2.0~\cite{parcollet2023lebenchmark} are used.
As we will apply these models on French pathological speech in further works, we targeted Wav2Vec2 models pre-trained exclusively on French speech.
Indeed, combining phones from multiple languages could complicate the analysis between healthy and pathological speech (meaning between typical and atypical French phones).
LeBenchmark models come in different architectures (6, 12, 24, or 48 hidden layers) as well as with different pre-training corpora sizes.
In this work, we will compare results obtained with models containing 6, 12, or 24 hidden layers, respectively referred to as \textit{light}, \textit{base}, and \textit{large}, as well as being pre-trained on 3k or 14k hours of French speech.
Wav2Vec2 works with \qty{25}{\ms}-windows, which approximately start every \qty{20}{\ms}, which implies an overlap of about \qty{5}{\ms}.
In order to emulate the CNN \qty{120}{\ms}-window, the input of Wav2Vec2 will consist here of audio extracts of six padded-windows of \qty{25}{\ms}.
Such an architecture gives us an approximate \qty{127}{\ms}-window, which is very close to the length of the context used for the CNN.

The output of either the CNN or the Wav2Vec2 model, once flattened, are passed through three 1024-dimension fully connected layers, dedicated to the phone classification task.
The size of the flattening layer depends on the output size of the encoder used (CNN or Wav2Vec2, and the size of Wav2Vec2).
Then, the output phone is chosen using a softmax on the 32 output values.

Datasets used for the training phase were randomly split into two phone-balanced subset: \textit{train} (90\% of data) and \textit{validation} (10\% left).
All of our models were fine-tuned using the SpeechBrain toolkit for 15 epochs, with a training time of approximately one day on a Tesla P100 GPU.
The model presenting the best phone error rate on the validation set was chosen for inference on test datasets.
The classifier uses an Adadelta optimizer with an initial learning rate of 0.9, to improve a cross-entropy loss applied on the phone classification.
The Wav2Vec2 architecture -- when fine-tuned -- relies on an Adam optimizer with an initial learning rate of $1.10^{-4}$.
Source code containing SpeechBrain recipes is made available on a Github repository\footnote{\scriptsize{\url{github.com/MaloMn/wav2vec2-phone-classification}}}.

\section{Experimental results}\label{sec:experiments}

\subsection{Comparison of Wav2Vec2 models}\label{subsec:wav2vec2}

Experiments were run to investigate the impact of the following factors:
\begin{inparaenum}[(1)]
  \item \label{item:fine-tuning} fine-tuning Wav2Vec2, 
  \item \label{item:pre-training} pre-training datasets,
  \item \label{item:model-size} model size, and
  \item \label{item:ft-data}fine-tuning datasets.
\end{inparaenum}
\Cref{tab:accuracies} sums up accuracies obtained for each fine-tuned model.
Given that the phone distribution is not balanced in C2SI datasets, accuracies are balanced by phone.
Mathematically, it is the average of accuracies obtained for each phone.
Thus, even when not specified, accuracies are systematically balanced by phone.

\begin{table}[t]
  \centering
  \caption[Phone-balanced accuracies (in \%) obtained on each test dataset.]{Phone-balanced accuracies (in \%) obtained on each test dataset.
  Confidence intervals were computed with the bootstrapping approach\protect\footnotemark.
  Results for the CNN are taken from \cite{abderrazek2023} and did not include confidence intervals.
  Best results are highlighted in bold.
  *The CNN model was trained from scratch using BREF.}
  \scalebox{0.71}{
    \begin{tabular}{p{0.27\linewidth}P{0.19\linewidth}P{0.2\linewidth}P{0.23\linewidth}P{0.23\linewidth}}
      \toprule
      \multirow{2}{*}{Model} & Fine-tuning datasets & \multirow{2}{*}{BREF-Int ↑} & C2SI-LEC ↑ (HC~speakers) & C2SI-DAP ↑ (HC~speakers) \\
      \midrule
      CNN, \textit{Baseline} & BREF* & 81.4 & 72.2 & 69.2 \\
      \midrule
      14k-large Frozen & BREF & 83.5$\pm$0.2 & 66.9$\pm$0.6 & 66.9$\pm$0.4 \\
      14k-large & BREF & 87.6$\pm$0.2 & 70.2$\pm$0.6 & 70.6$\pm$0.4 \\
      14k-light & BREF & 81.8$\pm$0.2 & 57.0$\pm$0.6 & 57.3$\pm$0.4 \\
      \midrule
      3k-large & BREF & \textbf{88.3}$\pm$\textbf{0.2} & 70.6$\pm$0.6 & 71.3$\pm$0.4 \\
      3k-base & BREF & 84.5$\pm$0.2 & 48.1$\pm$0.6 & 50.1$\pm${0.4} \\
      \midrule
      14k-large & BREF, CP & 87.4$\pm$0.2 & 72.1$\pm$0.5 & 73.3$\pm$0.4 \\
      14k-light & BREF, CP & 82.9$\pm$0.3 & 64.1$\pm$0.6 & 63.7$\pm$0.4 \\
      3k-large & BREF, CP & \textbf{88.3}$\pm$\textbf{0.2} & \textbf{72.6}$\pm$\textbf{0.6} & \textbf{73.9}$\pm$\textbf{0.4} \\
      3k-base & BREF, CP & 84.9$\pm$0.2 & 61.4$\pm$0.6 & 62.5$\pm$0.4 \\
      \bottomrule
    \end{tabular}
  }
  \label{tab:accuracies}
\end{table}

\footnotetext{\scriptsize{Ferrer, L. and Riera, P. Confidence Intervals for evaluation in machine learning [Computer software]. \url{https://github.com/luferrer/ConfidenceIntervals}}}

To analyse the impact of fine-tuning Wav2Vec2 (\ref{item:fine-tuning}), two \textit{14k-large} models were used, only one of which has been fine-tuned.
By comparing both models \textit{14k-large Frozen} and \textit{14k-large}, we can see that the fine-tuning step significantly improves accuracies on all three test datasets (confidence intervals are not overlapping themselves).
Unsurprisingly, fine-tuning Wav2Vec2 models on similar datasets is beneficial regarding accuracy.

Then, the impact of the pre-training datasets (\ref{item:pre-training}) was measured by using LeBenchmark models that had different sets of pre-training data.
We relied on both 3k and 14k model families, which are pre-trained on respectively 3,000 and 14,000 hours of French speech.
They include read, acted, spontaneous, and professional French speech, with standard and accentuated French.
Both of these datasets vary greatly in terms of accent variety.
Indeed, the 3k pre-training dataset mainly contains audiobooks and French radio programmes, with less than 1\% of African-accented speech that we are aware of -- included audiobooks and radio programmes do not provide information on the accents of their speakers.
However, the 14k pre-training datasets contain around 4,700 hours of speech from the European Parliament, which features at least three additional French accents: Belgian, Swiss, and Aosta Italian, as well as more negligible amounts of African radio, comprising Malian and Nigerian accents.
By comparing models \textit{14k-large} and \textit{3k-large}, our results underline that adding linguistic variability through various French accents does not offer significant improvements regarding classification accuracies.

Regarding model size (\ref{item:model-size}), multiple LeBenchmark model sizes (\ref{item:model-size}) were experimented : \textit{light}, \textit{base} and \textit{large}, involving 6, 12, and 24 hidden layers respectively.
Results show that using a \textit{large} model (with 24 hidden layers) offers significantly better accuracies than smaller models (with 6 and 12 hidden layers).
Unfortunately, as LeBenchmark offers neither a \textit{14k-base} model nor a \textit{3k-light} model, we cannot compare the results of the \textit{3k-base} and \textit{14k-light} models.
Nevertheless, our results show that smaller models have weaker generalization power than larger models on both unseen C2SI datasets.

Finally, to analyse the impact of fine-tuning datasets on transformer-based models (\ref{item:ft-data}), we add a new fine-tuning dataset -- the Common Phone (CP) corpus -- compared with~\cite{abderrazek2023}, in which only the BREF dataset was involved.
All the initial models mentioned above have separately been fine-tuned on a combination of the BREF and Common Phone corpora (\textit{BREF,CP} in \Cref{tab:accuracies}).
According to our results, adding read speech from other corpora in the fine-tuning stage significantly improves the generalization to both C2SI datasets on all our fine-tuned models (for instance, an absolute 12.4\% improvement is achieved by the \textit{3k-base} model trained on BREF and CP, compared with the \textit{3k-base} model trained on BREF only, on the C2SI-DAP dataset).
Accuracies are also similar on the BREF-Int dataset: all confidence intervals overlap each other, except for the \textit{14k-light} model, which is significantly better, once fine-tuned on the combination of BREF and Common Phone.
It therefore seems worthwhile to include other datasets in the fine-tuning stage, even if they do not represent a significant percentage of the training data.

\subsection{Wav2Vec2 and CNN}\label{subsec:cnn}
\begin{figure}[b]
  \vspace{-1em}
  \centering
  \subfloat[BREF-Int]{\includegraphics[width=0.23\textwidth]{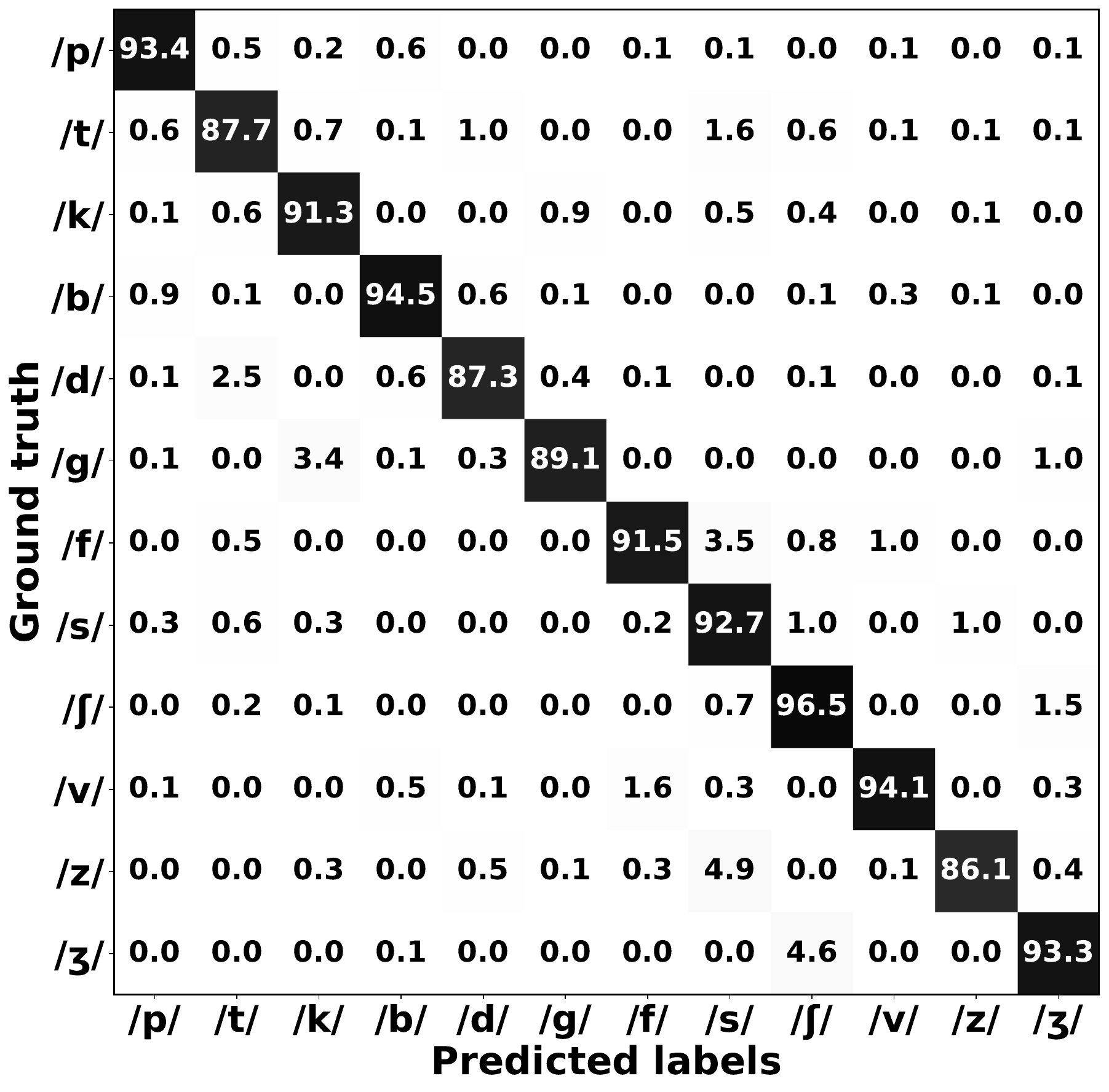}\label{fig:bref-obstruent}}
  \hfill
  \subfloat[C2SI-LEC]{\includegraphics[width=0.23\textwidth]{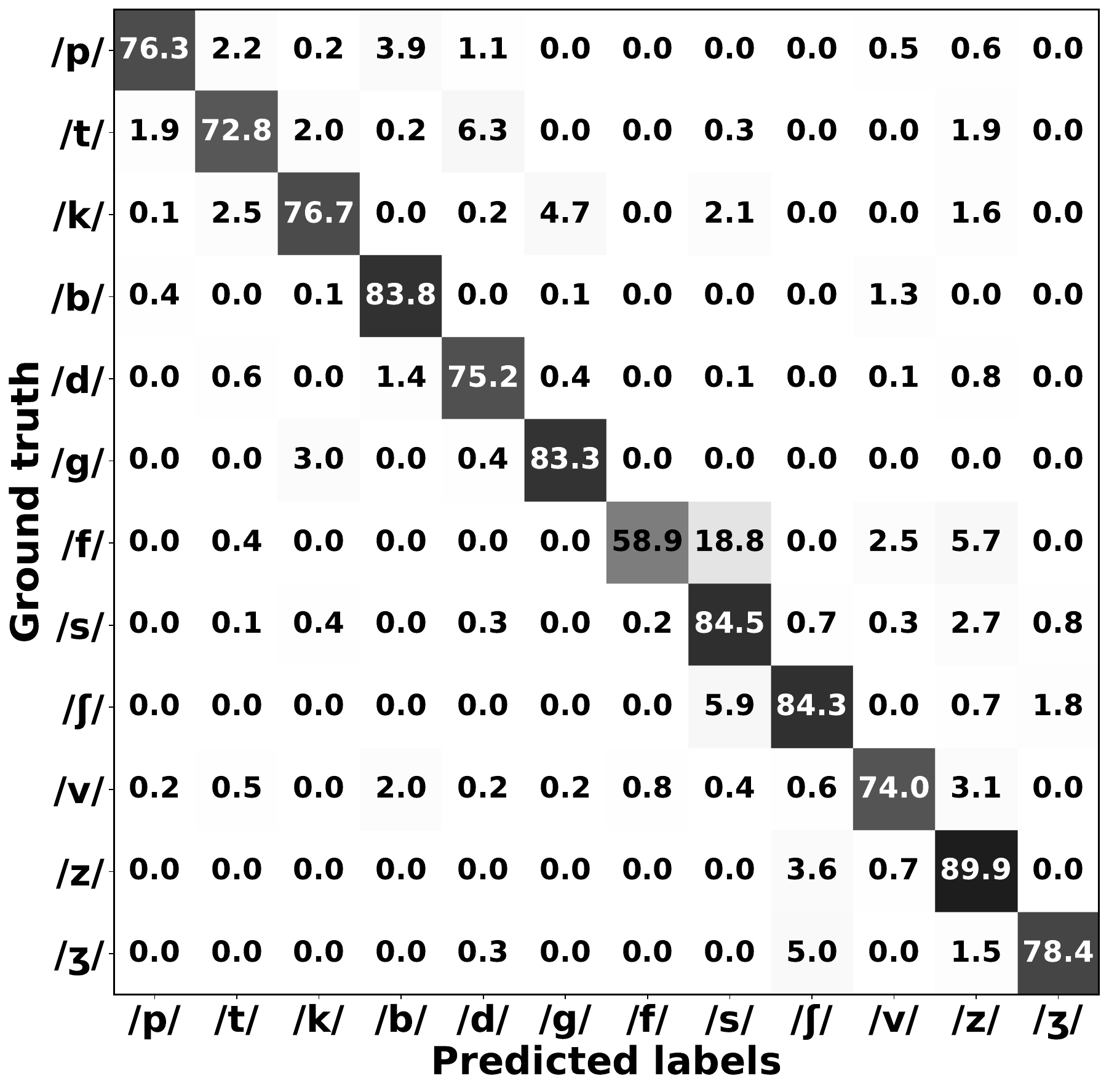}\label{fig:c2si-obstruent}}  
  \caption{Confusion matrices of obstruent phones, on BREF-Int and C2SI-LEC (HC speakers only) datasets respectively.}  
\end{figure}
The best Wav2Vec2 model -- \textit{3k-large} fine-tuned on BREF and CP, still according to results shown in \cref{tab:accuracies}, improves by an absolute 6.9\% the classification accuracy on BREF-Int compared with the CNN (88.3\% versus 81.4\%).
On C2SI-LEC, the difference is not significant (72.6\% versus 72.2\%).
Comparatively, we note a significant improvement of 4.7\% on C2SI-DAP (73.9\% versus 69.2\%).
However, by comparing the number of parameters of these models, we can point out that the best Wav2Vec2 model has 330 million of parameters, against 10 million for the CNN.
Furthermore, smaller Wav2Vec2 models \textit{base} and \textit{light} -- with respectively 90 and 26 million parameters -- do not generalize as well as the CNN on C2SI datasets.
This gap in parameter count is significant, but both model types do not have the same architecture, and a comparison based only on parameter count is necessarily biased.
Nevertheless, this size difference implies that the carbon footprint of our inferences will inevitably be higher when using Wav2Vec2 models.

\begin{figure}[t]
  \centering
  \subfloat[BREF-Int]{\includegraphics[width=0.23\textwidth]{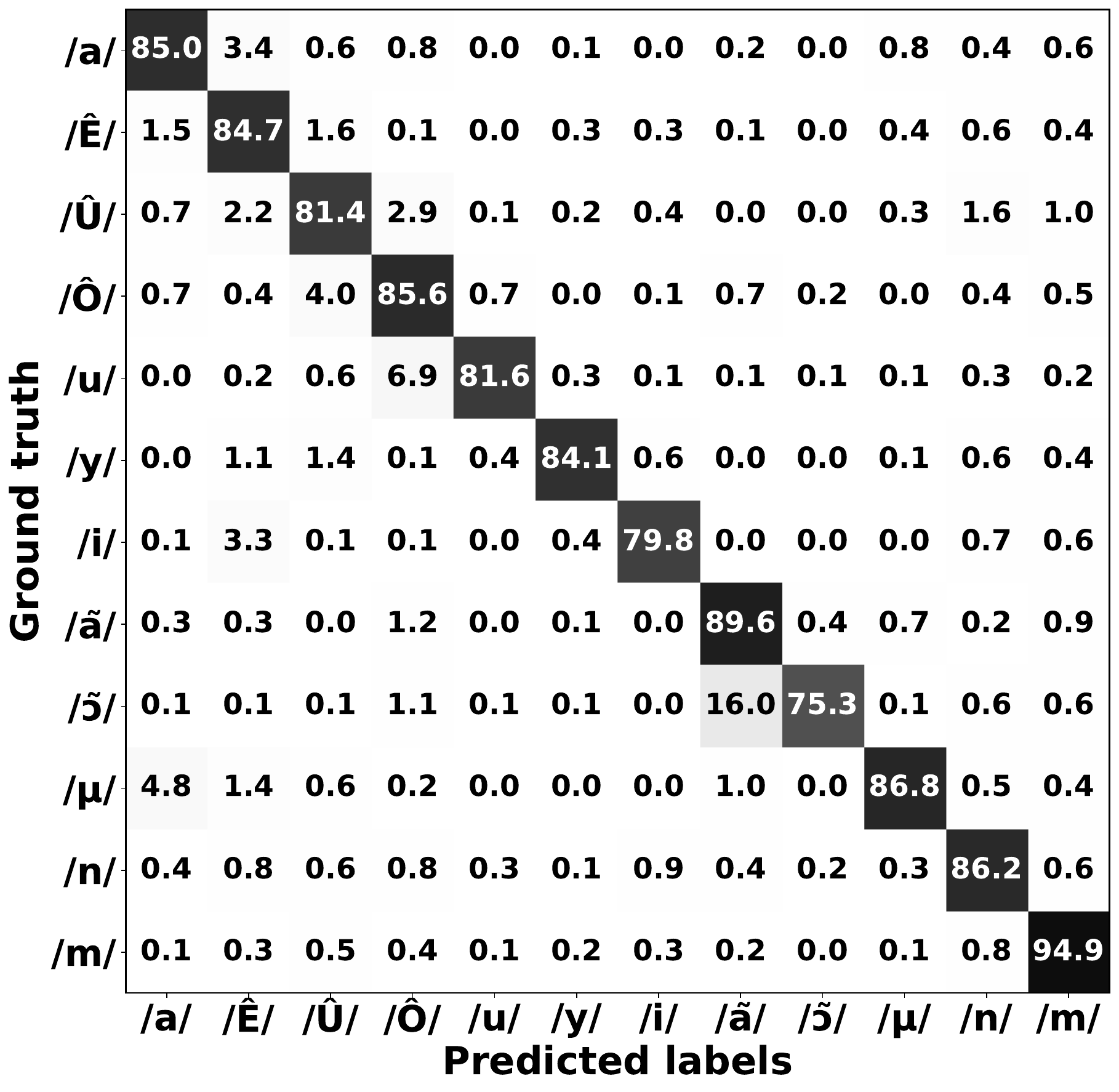}\label{fig:bref-oral}}
  \hfill
  \subfloat[C2SI-LEC]{\includegraphics[width=0.23\textwidth]{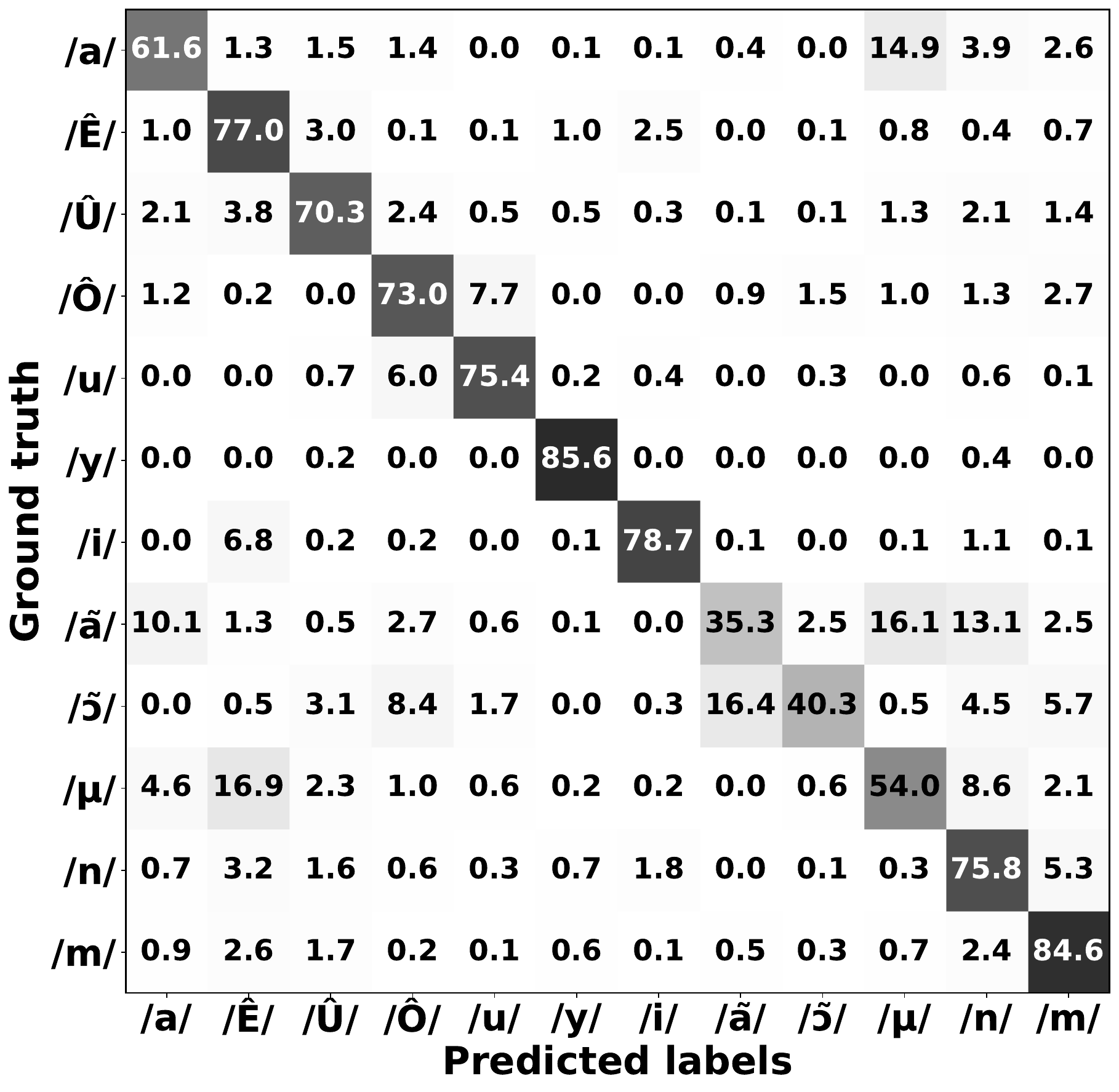}\label{fig:c2si-oral}}
  \caption{Confusion matrices of oral/nasal phones, on BREF-Int and C2SI-LEC (HC speakers only) datasets respectively.}
\end{figure}

To ensure that our models do not overfit on certain phones, and that confusions between phones remain explainable, confusion matrices were generated on BREF-Int and C2SI-LEC, using the \textit{3k-large} model fine-tuned on BREF and Common Phone.
\Cref{fig:bref-obstruent,fig:bref-oral,fig:c2si-obstruent,fig:c2si-oral} show specific parts of these matrices, dedicated to obstruent and oral/nasal phones.
Restraining our analysis to these phones makes a direct comparison with the analysis done in~\cite{abderrazek2023} possible.
The comparison between previously obtained results and current results show that Wav2Vec reduces the observed confusions in most cases.

Regarding obstruent phones (\cref{fig:bref-obstruent,fig:c2si-obstruent}), /\textyogh/ was confused with /\textesh/ in 9\% of cases on BREF-Int in the previous work, against 4.6\% now.
Confusion between both of these phones is therefore still present, but happens in half the cases compared with previous results.
This reduction can also be found on the C2SI dataset: whereas /p/ was previously confused with /t/ in 9.3\% of cases, this is only the case 2.2\% of the time with our architecture.
We also find the same causes for important confusions: either the loss of the place of articulation (acuteness -- /f/ → /s/, compactness -- /\textesh/ → /s/), or confusion due to the voicing feature (/t/~→~/d/).

Regarding oral and nasal phones, we also get strong confusions linked to oral vowels and nasal consonants on C2SI-LEC.
Where /\~a/ and /a/ were confused in 11.5\% of cases, they still are in 10.1\% of cases.
/\~a/ is also more confused now with /n/ (13.1\%, against 8.4\% previously) but is less confused with /m/ (2.5\%, against 6.2\% previously).
These confusions, previously explained by the change of speakers' accent (Parisians for BREF-Int, and from the Toulouse area for C2SI), can also be found here, when using another type of model architecture.
Our results therefore support those obtained previously, and show that the data used during fine-tuning and the domain differences between datasets remain a non-negligible problem for this type of architecture.
On BREF-Int, we also find strong confusions between /\~a/ and /\~c/, with a mutual confusion of 16.4\%, albeit less intense than when using a CNN -- 20.6\%.
These results are important because they show that our model is also sensitive to atypical pronunciations (such as a regional accent), which is desirable when analysing pathological speech.

\subsection{Application to pathological speech}\label{subsec:patho}

\begin{figure}[t]
  \centering
  \subfloat[DES-Severity]{\includegraphics[width=0.235\textwidth]{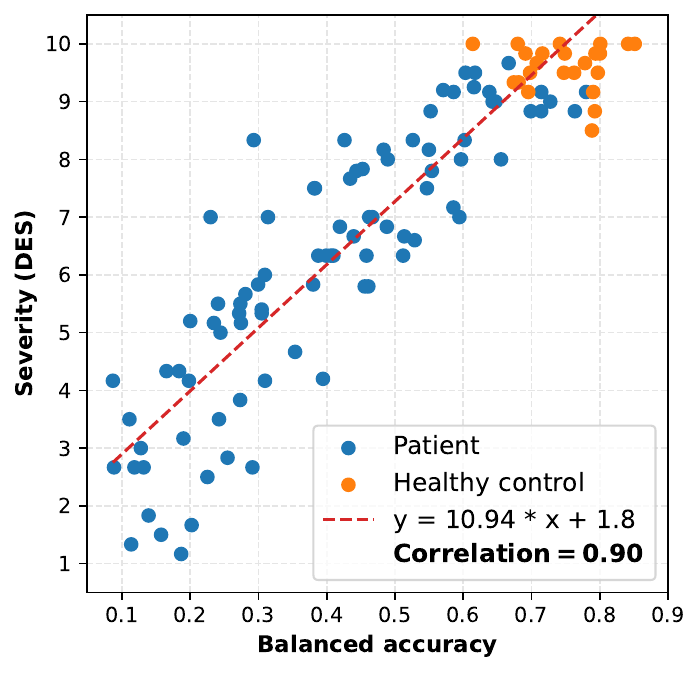}\label{fig:severity}}
  \hfill
  \subfloat[DES-Intelligibility]{\includegraphics[width=0.235\textwidth]{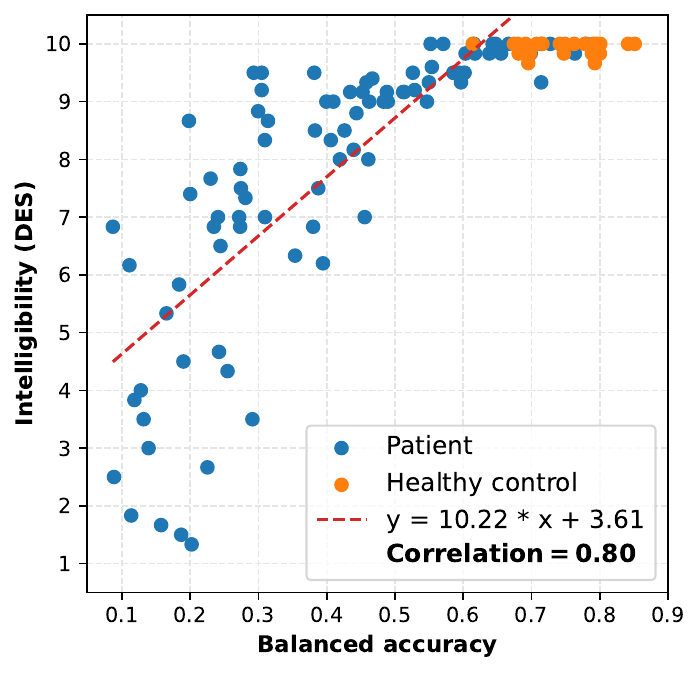}\label{fig:intelligibility}}
  \caption{Scatter plots of C2SI patients and HC on perceptual evaluations by experts and balanced accuracies obtained.}  
\end{figure}

Having demonstrated the robustness of our model and its ability to generalize well to other datasets, we will now see whether its computed accuracies can correlate positively with the assessments of the six C2SI experts.
The severity and intelligibility scores given by the experts on the image description task (DES) for a speech record were averaged and then compared to classification accuracies.
\Cref{fig:severity,fig:intelligibility} show scatter plots comparing perceptual scores to classification accuracies of phones in recordings of patients and HC from the C2SI corpus.
Regression lines are also plotted.
High Pearson correlation values -- 0.90 for severity and 0.80 for intelligibility -- confirm that perceptual measures can be estimated using phone-balanced classification accuracies with our \textit{3k-large} model fine-tuned on BREF and Common Phone.
These correlation values are similar to the ones obtained using the CNN~\cite{abderrazek2023}, where balanced accuracies correlated at 0.91 and 0.81 with severity and intelligibility scores respectively.

The less-intense correlation with perceptual intelligibility can be explained by the fact that experts are known to be biased on evaluating intelligibility, by underestimating the degree of speech disorders. 
Indeed, when a task involves using expected French words (for instance, using the word ``bateau'' to describe an image showing a boat), experts tend to understand them better.
This was one of the reasons for creating a pseudo-word reading task in \cite{woisard2021c2si,lalain2020}.
Furthermore, experts have generally less trouble understanding patients.
These results confirm that a speech representation based on a Wav2Vec2-based model is well suited for a phone-wise analysis of pathological speech.
\section{Conclusion}\label{sec:conclusion}
In this work, we showed that a Wav2Vec2-based model outperforms a CNN model on a phone classification task, while preserving certain linguistic specificities, such as regional accents.
Our results validate not only the efficacy of our approach based on Wav2Vec2, but also highlights the importance of the choice of the model architecture as well as the diversity of fine-tuning datasets for optimal performance.
Future works include applying the NCD concept developed in~\cite{abderrazek2023} to analyse hidden layers of our fine-tuned architecture, and analyse how this new model influences the detection of phonetic features, which are crucial for interpreting predicted phones.
In turn, this interpretability is needed to have an objective analysis of a patient's speech, which would improve re-habilitation techniques implemented by clinicians.

\bibliographystyle{IEEEtran}
\bibliography{mybib}

\begin{thebibliography}{10}
\providecommand{\url}[1]{#1}
\csname url@samestyle\endcsname
\providecommand{\newblock}{\relax}
\providecommand{\bibinfo}[2]{#2}
\providecommand{\BIBentrySTDinterwordspacing}{\spaceskip=0pt\relax}
\providecommand{\BIBentryALTinterwordstretchfactor}{4}
\providecommand{\BIBentryALTinterwordspacing}{\spaceskip=\fontdimen2\font plus
\BIBentryALTinterwordstretchfactor\fontdimen3\font minus
  \fontdimen4\font\relax}
\providecommand{\BIBforeignlanguage}[2]{{%
\expandafter\ifx\csname l@#1\endcsname\relax
\typeout{** WARNING: IEEEtran.bst: No hyphenation pattern has been}%
\typeout{** loaded for the language `#1'. Using the pattern for}%
\typeout{** the default language instead.}%
\else
\language=\csname l@#1\endcsname
\fi
#2}}
\providecommand{\BIBdecl}{\relax}
\BIBdecl

\bibitem{woisard2021c2si}
C.~Ast{\'e}sano, M.~Balaguer, J.~Farinas, C.~Fredouille, P.~Gaillard, A.~Ghio,
  I.~Laaridh, M.~Lalain, B.~Lepage, J.~Mauclair, O.~Nocaudie, J.~Pinquier,
  O.~Pont, G.~Pouchoulin, M.~Puech, D.~Robert, E.~Sicard, and V.~Woisard,
  ``Carcinologic speech severity index project: A database of speech disorder
  productions to assess quality of life related to speech after cancer,'' in
  \emph{Proceedings of the Eleventh International Conference on Language
  Resources and Evaluation ({LREC} 2018)}.\hskip 1em plus 0.5em minus
  0.4em\relax Miyazaki, Japan: European Language Resources Association (ELRA),
  May 2018.

\bibitem{maier2010}
A.~Maier, H.~Tino, F.~Stelzle, E.~Noeth, E.~Nkenke, R.~Frank, S.~Anne, and
  M.~Schuster, ``Automatic speech recognition systems for the evaluation of
  voice and speech disorders in head and neck cancer,'' \emph{EURASIP Journal
  on Audio, Speech, and Music Processing}, vol. 2010, 01 2010.

\bibitem{middag2014}
C.~Middag, R.~Clapham, R.~{van Son}, and J.-P. Martens, ``Robust automatic
  intelligibility assessment techniques evaluated on speakers treated for head
  and neck cancer,'' \emph{Computer Speech \& Language}, vol.~28, no.~2, pp.
  467--482, 2014.

\bibitem{laaridh18b}
I.~Laaridh, C.~Fredouille, A.~Ghio, M.~Lalain, and V.~Woisard, ``{Automatic
  Evaluation of Speech Intelligibility Based on I-vectors in the Context of
  Head and Neck Cancers},'' in \emph{Proc. Interspeech 2018}, 2018, pp.
  2943--2947.

\bibitem{vaysse2021}
R.~Vaysse, J.~Farinas, C.~Astésano, and R.~Andre-Obrecht, ``Automatic
  extraction of speech rhythm descriptors for speech intelligibility assessment
  in the context of head and neck cancers,'' 08 2021, pp. 1912--1916.

\bibitem{bin2019}
L.~Bin, M.~C. Kelley, D.~Aalto, and B.~V. Tucker, ``Automatic speech
  intelligibility scoring of head and neck cancer patients with deep neural
  networks,'' in \emph{International Congress of Phonetic Sciences (ICPHs’
  19), Melbourne, Australia}, 2019, pp. 3016--3020.

\bibitem{quintas2023}
S.~Quintas, A.~Abad, J.~Mauclair, V.~Woisard, and J.~Pinquier, ``Towards
  reducing patient effort for the automatic prediction of speech
  intelligibility in head and neck cancers,'' in \emph{ICASSP 2023 - 2023 IEEE
  International Conference on Acoustics, Speech and Signal Processing
  (ICASSP)}, 2023, pp. 1--5.

\bibitem{abderrazek2023}
S.~Abderrazek, C.~Fredouille, A.~Ghio, M.~Lalain, C.~Meunier, and V.~Woisard,
  ``Interpreting deep representations of phonetic features via neuro-based
  concept detector: Application to speech disorders due to head and neck
  cancer,'' \emph{IEEE/ACM Transactions on Audio, Speech, and Language
  Processing}, vol.~31, pp. 200--214, 2023.

\bibitem{dieck22_interspeech}
T.~tom Dieck, P.~A. Pérez-Toro, T.~Arias, E.~Noeth, and P.~Klumpp, ``{Wav2vec
  behind the Scenes: How end2end Models learn Phonetics},'' in \emph{Proc.
  Interspeech 2022}, 2022, pp. 5130--5134.

\bibitem{baevski2020}
A.~Baevski, Y.~Zhou, A.~Mohamed, and M.~Auli, ``wav2vec 2.0: A framework for
  self-supervised learning of speech representations,'' in \emph{Advances in
  Neural Information Processing Systems}, vol.~33.\hskip 1em plus 0.5em minus
  0.4em\relax Curran Associates, Inc., 2020, pp. 12\,449--12\,460.

\bibitem{pasad2021}
A.~Pasad, J.-C. Chou, and K.~Livescu, ``Layer-wise analysis of a
  self-supervised speech representation model,'' in \emph{2021 IEEE Automatic
  Speech Recognition and Understanding Workshop (ASRU)}, 2021, pp. 914--921.

\bibitem{pasad2023}
A.~Pasad, B.~Shi, and K.~Livescu, ``Comparative layer-wise analysis of
  self-supervised speech models,'' in \emph{ICASSP 2023 - 2023 IEEE
  International Conference on Acoustics, Speech and Signal Processing
  (ICASSP)}, 2023, pp. 1--5.

\bibitem{hernandez22_interspeech}
A.~Hernandez, P.~A. Pérez-Toro, E.~Noeth, J.~R. Orozco-Arroyave, A.~Maier, and
  S.~H. Yang, ``{Cross-lingual Self-Supervised Speech Representations for
  Improved Dysarthric Speech Recognition},'' in \emph{Proc. Interspeech 2022},
  2022, pp. 51--55.

\bibitem{yeo23icassp}
E.~J. Yeo, K.~Choi, S.~Kim, and M.~Chung, ``Automatic severity classification
  of dysarthric speech by using self-supervised model with multi-task
  learning,'' in \emph{ICASSP 2023 - 2023 IEEE International Conference on
  Acoustics, Speech and Signal Processing (ICASSP)}, 2023, pp. 1--5.

\bibitem{javanmardi2024}
F.~Javanmardi, S.~R. Kadiri, and P.~Alku, ``Pre-trained models for detection
  and severity level classification of dysarthria from speech,'' \emph{Speech
  Communication}, p. 103047, 2024.

\bibitem{favaro2023}
A.~Favaro, Y.-T. Tsai, A.~Butala, T.~Thebaud, J.~Villalba, N.~Dehak, and
  L.~Moro-Velázquez, ``Interpretable speech features vs. dnn embeddings: What
  to use in the automatic assessment of parkinson’s disease in multi-lingual
  scenarios,'' \emph{Computers in Biology and Medicine}, vol. 166, p. 107559,
  2023.

\bibitem{ming2017}
M.~Tu, V.~Berisha, and J.~Liss, ``{Interpretable Objective Assessment of
  Dysarthric Speech Based on Deep Neural Networks},'' in \emph{Proc.
  Interspeech 2017}, 2017, pp. 1849--1853.

\bibitem{darley1969b}
F.~L. Darley, A.~E. Aronson, and J.~R. Brown, ``Clusters of deviant speech
  dimensions in the dysarthrias,'' \emph{Journal of Speech and Hearing
  Research}, vol.~12, no.~3, pp. 462--496, 1969.

\bibitem{xu2023}
L.~Xu, J.~Liss, and V.~Berisha, ``{Dysarthria detection based on a deep
  learning model with a clinically-interpretable layer},'' \emph{JASA Express
  Letters}, vol.~3, no.~1, p. 015201, 01 2023.

\bibitem{lundberg2017}
S.~M. Lundberg and S.-I. Lee, ``A unified approach to interpreting model
  predictions,'' in \emph{Advances in Neural Information Processing Systems},
  vol.~30.\hskip 1em plus 0.5em minus 0.4em\relax Curran Associates, Inc.,
  2017.

\bibitem{yeo23interspeech}
E.~J. Yeo, K.~Choi, S.~Kim, and M.~Chung, ``Speech intelligibility assessment
  of dysarthric speech by using goodness of pronunciation with uncertainty
  quantification,'' in \emph{Proc. INTERSPEECH 2023}, 2023, pp. 166--170.

\bibitem{lamel1991bref}
L.~F. Lamel, J.-L. Gauvain, M.~Esk{\'e}nazi \emph{et~al.}, ``Bref, a large
  vocabulary spoken corpus for french.'' \emph{Eurospeech’91, Italy},
  vol.~22, no.~28, p.~50, 1991.

\bibitem{klumpp2022cp}
P.~Klumpp, T.~Arias, P.~A. P{\'e}rez-Toro, E.~Noeth, and J.~Orozco-Arroyave,
  ``Common phone: A multilingual dataset for robust acoustic modelling,'' in
  \emph{Proceedings of the Thirteenth Language Resources and Evaluation
  Conference}.\hskip 1em plus 0.5em minus 0.4em\relax Marseille, France:
  European Language Resources Association, Jun. 2022, pp. 763--768.

\bibitem{parcollet2023lebenchmark}
T.~Parcollet, H.~Nguyen, S.~Evain, M.~{Zanon Boito}, A.~Pupier, S.~Mdhaffar,
  H.~Le, S.~Alisamir, N.~Tomashenko, M.~Dinarelli, S.~Zhang, A.~Allauzen,
  M.~Coavoux, Y.~Estève, M.~Rouvier, J.~Goulian, B.~Lecouteux, F.~Portet,
  S.~Rossato, F.~Ringeval, D.~Schwab, and L.~Besacier, ``Lebenchmark 2.0: A
  standardized, replicable and enhanced framework for self-supervised
  representations of french speech,'' \emph{Computer Speech \& Language},
  vol.~86, p. 101622, 2024.

\bibitem{lalain2020}
M.~Lalain, A.~Ghio, L.~Giusti, D.~Robert, C.~Fredouille, and V.~Woisard,
  ``Design and development of a speech intelligibility test based on
  pseudowords in french: Why and how?'' \emph{Journal of Speech, Language, and
  Hearing Research}, vol.~63, no.~7, pp. 2070--2083, 2020.

\end{thebibliography}

\end{document}